\documentstyle[aps,preprint]{revtex}
\begin{document}
\title{Maximal Bell's Inequality Violation for Non Maximal Entanglement }
\author{M. Kobayashi$^{1,2}$ ,\,\,  F. Khanna$^2$ ,\,\,A. Mann$^3$,\,\,
 M. Revzen$^{2,3}$ ,\,\, A. Santana$^4$ }
\address{$^1$ Department of Physics, Gifu University, Gifu, Japan.\\
 $^2$ Physics Department,  University of Alberta, Edmonton ,  Alberta, 
Canada T6G 2J1.\\
$^3$  Department of Physics, Technion - Israel Institute of Technology ,
Haifa  32000, Israel. \\
$^4$  Instituto de F$\grave i$sica, Universidade Federal da Bahia, Salvador, 
Bahia, Brazil. } 
\maketitle

\begin{abstract}
 Bell's inequality violation (BIQV) for correlations of polarization is studied
for a {\it product} state of two two-mode squeezed vacuum (TMSV) states. The
violation allowed is shown to attain its maximal limit for all values of
the squeezing parameter, $\zeta$.  We show via an explicit example that 
a state whose entanglement is not maximal allow maximal BIQV. 
The Wigner function of the state is non negative and the average value of 
either polarization is nil.

\end{abstract}
\bigskip
\noindent ${\rm PACS}:\;2.0; 03.65.{\rm Bz}$
\bigskip

Two mode squeezed vacuum (TMSV) states generated via nondegenerate parametric
amplifiers are of interest, following the pioneering study of Grangier et al.
\cite{grangier} that exhibited Bell's inequality violations (BIQV) 
\cite{peres},\cite{shimony} via 
intensity correlation measurements \cite{walls} for such states. 
(The Bell inequality we
 consider in this paper is the so called CHSH inequality, \cite{clauser}.)
These BIQV studies, aside from their intrinsic interest, are valuable for 
elucidation of the applicability
of Bell's inequalities for continuous variables problems in general, and to
the original Einstein, Podolsky and Rosen (EPR) \cite{epr} state in 
particular. The 
latter's pertinency to BIQV was discussed by Bell \cite{bell}. The
reason for this general interest is associated with the fact that the Wigner
 function  \cite{wigner} for the EPR state (and, in general, for the 
TMSV state) is
non-negative and thus has some attributes of
a probability density in phase space, i.e. the problem relates to the 
fundamental question as to whether the quantum behaviour of these states
can be underpinned with a theory based on local hidden variables.  
Bell \cite{bell} considered the problem and discussed correlations
among the particles' positions. He suggested  \cite{bell} that such 
correlations within the EPR state (whose Wigner function is non
-negative) will not allow BIQV.
 Wodkiewicz and Banaszek \cite
{wodkiewicz} noted that, intrinsically, \cite{royer,englert,knight}, it 
is the {\it parity} 
correlation that is expressed by the Wigner function.
They \cite{wodkiewicz} then showed that the TMVS allow BIQV via 
{\it parity} correlations measurements. This analysis was greatly advanced
by explicit definitions of ``rotation'' in parity space in \cite{chen,gour},
where it is shown that BIQV can achieve its Cirelson's limit \cite{cirelson} 
for maximal entanglement, i.e. $\zeta \rightarrow \infty$,
where $\zeta$ is the (positive) squeezing parameter. The above underscores 
the importance of specifying the operators
(i.e., the dynamical variables) involved in the definition of the Bell 
operator \cite{braunstein}, whose correlation values are bounded by Bell's 
inequality.
In the present work we consider {\it polarization} as our dynamical variable 
(the representative operator is given below -  above Eq. (17)) . We 
show 
that the state under study: $|\zeta \rangle$, a {\it product} of two TMSV 
states, whose Wigner function is non-negative, exhibits the 
remarkable property of allowing maximal BIQV , i.e. attaining the Cirelson 
limit,   for all values of the squeezing parameter 
$\zeta$.\\ 
The state under study, $|\zeta \rangle$, is given by (\cite{faqir})
\begin{equation}
 |\zeta \rangle = exp(\zeta  K_{x }) |0\rangle  ;    \label{first}
 \end{equation}
\begin{equation}
  K_x\;  =\; a_{+} ^{+}b_{+}^{+}  +  a_{-}^{+}b_{-}^
{+}   - a_{+}b_{+} -  a_{-}b_{-}.  \label{dir1}
\end{equation}
Here the $+/-$ subscripts denote the polarization  relative to some chosen 
axis, common to A and B, with A and B labeling the two different 
channels
toward  which the two beams head - the operators $a$ and $b$ refer to the 
respective 
channels: e.g. $a_{+} ^{+}$ designates the creation operator for horizontally,
i.e., `` x polarized'' \cite{eberly} photon headed into the A channel, etc. 
$|\zeta \rangle$ is a product of two two-mode squeezed vacua - one pertaining  to x- polarized and one to y- polarized. The Wigner function $W(\alpha_{A},
\alpha_{B})$
for the state is thus, likewise, a product of two functions each of the form
 \cite{mandel} ($\alpha = q + ip$;  we delete the polarization 
subscripts to reduce notational cluttering)
\begin{equation}
  W(\alpha_{A},\alpha_{B}) = {4 \over \pi^{2}}exp[-2cosh(2\zeta)(|\alpha_{A}|
^{2} + |\alpha_{B}|^{2}) +2sinh(2\zeta)(\alpha_{A}\alpha_{B} + c.c.)].
\end{equation}
The total Wigner function, being a product of two such gaussian functions,
is clearly nonnegative. 
 
We further recall the symmetry operator \cite{faqir}, 
 \begin{equation}
 K_{0}\;  =  \;i[a_{+}^{+} a_{-}\;-\; a_{-}^{+}a_{+}\; +\; 
 b_{+}^{+} b_{-}\; -\;  b_{-}^{+}b_{+}].   \label{new1}
\end{equation}
 Since 
 \begin{equation}
[K_0, K_x]_{-}    =   0,
 \end{equation}
 and
 \begin{equation}
 K_0 | 0\rangle     =   0,
 \end{equation}
 we have
\begin{equation}
 |\zeta^{'}   \rangle \;\;{\equiv }\;\;  exp( -i\delta K_0) 
|\zeta \rangle\; \;  =\;\;   |\zeta \rangle , 
 \end{equation}
  i.e.  $ exp(-i\delta K_0)$  is a symmetry operator: it leaves the state
invariant. Note that $K_0$ is made of , additively, two parts

\begin{equation}
K_{0}^{A} \;=\; i [ a_{+}^{+} a_{-}  -  a_{-}^{+} a_{+}] ,\;\;\;and \;\;
K_{0}^{B} \;=\; i [b_{+}^{+} b_{-}  -  b_{-}^{+} b_{+}].
\end{equation}
Each of these parts acts on a distinct channel - which we take to be at a 
different locale, but
\begin{equation}
 K_{0}  =  K_{0}^{A}  +  K_{0}^{B},
\label{sym}
\end{equation}
is a symmetry operator for the system as a whole. On the other hand, the 
operator \, 
 $exp(i\delta _{A} K_{0}^{A})$ \,\, is a symmetry {\it breaking} 
 operator - it breaks the \,\,$K_{0}$ \,\, symmetry. This operator does 
{\it not} commute with \,\,$K_{x}$ .

Rotating the polarizations, ~\cite{eberly}, i.e., e.g., $ a_{+}^{+}
 \rightarrow a_{\delta_{A}}^{+}$, is accomplished via 
\begin{eqnarray*}
a_{\alpha}^{+}  (\delta _{A}) &=& \exp^{i\delta _{A} K_{0}^{A}} a_{\alpha}^{+}\exp^{-i\delta _{A} K_{0}^{A}}, \,\, \, \alpha  = \pm  ,\\
b_{\beta}(\delta _{B})&=&\exp^{i\delta _{B} K_{0}^{B}}b_{\beta}\exp^{-i\delta _{B}K_{0}^{B}},\,\,\, \beta =\pm, 
\end{eqnarray*}
and their hermitian adjoints. We shall study the correlations between 
polarizations of the two channels. To this end our normalizer (i.e., 
reference correlation) is,
\begin{equation}
C(\;A,\;B) = \langle \zeta|(a_{+}^{+} a_{+}  -  a_{-}^{+} a_{-})
(b_{+}^{+} b_{+}  -  b_{-}^{+} b_{-})|\zeta \rangle.
\end{equation}
It is shown in the appendix of \cite{faqir} that the correlation function
$$
C^{\alpha \beta}(\delta _{A}, \delta _{B}) = \langle \zeta |
\exp ^{i\delta _{A} K_{0}^{A} + i\delta _{B}K_{0}^{B}}a_{\alpha }^{+}
 a_{\alpha }b_{\beta }^{+}b_{\beta  } \exp^{-i\delta _{A}K_{0}^{A} 
-i\delta _{B}K_{0}^{B}}|\zeta \rangle,\;\;\; \alpha, \beta = \pm,
$$ 
is a function of $\delta\;=\; \delta_{A}\; -\; \delta_{B}.$  Hence the 
operator,  $ exp(-i\delta K_0)$  implies 
\begin{equation}
C(A(\delta), B(\delta)) = \langle \zeta|(a_{\delta}^{+} a_{\delta}  - 
 a_{\bar\delta}^{+} a_{\bar\delta})
(b_{\delta}^{+} b_{\delta}  -  b_{\bar\delta}^{+} b_{\bar\delta})|\zeta \rangle
\;\;= \;\;C(\;A(0),\;B(0))\;\equiv \;C(\;A,\;B),
\label{diff} 
\end{equation}
with $\bar\delta = \delta \pm \pi/2$ (cf. \cite{eberly}). The polarization
along $\bar\delta$ is orthogonal to the one along $\delta$. i.e. the 
correlation is invariant under equal (common) rotation of the two channels.
Direct calculation gives
\begin{equation}
C(A(\delta), B(\delta'))\;=\;C^{++}(\delta, \delta') + C^{--}(\delta, \delta')
 - C^{+-}(\delta, \delta') - C^{-+}(\delta, \delta'),
\end{equation}
thus,

\begin{equation}
C(\;A,\;B) =\; 2cosh^{2}\zeta\; sinh^{2}\zeta.
\end{equation}
Similarly, direct calculation yields  
\begin{equation}
C(A(\delta), B(\delta')) = C(\;A,\;B)cos2(\delta - \delta').
\end{equation}
Define the normalized expectation value of the polarization correlation 
by
\begin{equation}
E(\delta,\delta') = {C(A(\delta), B(\delta')) \over \sqrt{
C(A(\delta), B(\delta))\;C(A(\delta'), B(\delta'))}} = cos2(\delta - \delta').
\label{E}
\end{equation}
Our expression for the Bell inequality \cite{walls} is, thus 
\begin{equation}
|E(\delta_{A},\delta_{B})  +  E(\delta  _{A}, \delta  _{B}^{'})   +   E(\delta  _{A}^{'}, \delta  _{B})  -  E(\delta  _{A}^{'}, \delta  _{B}^{'})|    \leq  2.
\end{equation}
Using Eq.(\ref{E}), the left hand side can be written as
\begin{equation}
 |cos 2(\delta _{A} - \delta _{B})  +  cos 2(\delta _{A} - \delta _{B}^{'}) 
 +  cos 2(\delta _{A}^{'} - \delta _{B}) - cos 2(\delta _{A}^{'} -
\delta _{B}^{'})| \;\;  \leq  2.
\end{equation}
Clearly the choice  $\delta_{A}\; =\; 0,\; \delta_{A}'\; =\; \pi/4,\; 
\delta_{B}\; = \;\pi/8,\; \delta_{B}'\; =\; -\pi/8,$  gives for the LHS 
$2\sqrt2$,
 i.e. {\it maximal} violation of the inequality. This result is independent 
of $\zeta$. We recall that for each
polarization, $\pm$, the entanglement of the state is maximized only for
$\zeta \rightarrow \infty$  \cite{wodkiewicz},\cite{chen},\cite{gour}, i.e., 
not only
is the state under study a product of two states, but also both 
component states are {\it not} maximally entangled (for $\zeta\; <\; 
\infty$). However, BIQV 
 discussed in these references are for {\it parity} as the relevant 
dynamical
variable. In our study the relevant dynamical variable is 
the {\it polarization} in each channel, viz: $(a_{+}^{+} a_{+}  -  
a_{-}^{+} a_{-})$. For this 
variable the amount of BIQV can be estimated by evaluating its average 
value:
\begin{equation}
\langle \zeta| (a_{+}^{+} a_{+}  -  a_{-}^{+} a_{-})|\zeta \rangle = 0,
\end{equation}

which holds for all values of  $\zeta$ for the pure state $|\zeta \rangle$
  (Eq.(1)), and for
both channels variables. This value, (nil) in Eq. (18),  
 is analogous to the two spin state case,
such as, e.g. (the numerical superscripts label the particles),  
$$|\psi \rangle = \alpha |\uparrow^{1}\uparrow^{2} \rangle\;\;+\;\;
\beta|\downarrow^{1}\downarrow^{2}\rangle,$$
where  $\langle \psi|\sigma_{z}^{(1)}|\psi\rangle \;\;=\;\;  \langle \psi|
\sigma_{z}^{(2)}|\psi\rangle\;\;=\;\;0$ implies allowance of maximal BIQV
{\it and}, in the spin case,  $|\alpha|\, = \, |\beta|$ i.e. maximal 
entanglement.

 For comparison, we evaluate (just for one polarization, +) the expectation 
value of the  
 {\it parity} operator \cite{chen},\cite{gour} as the relevant 
dynamical variable,
\begin{equation}
\langle \zeta|\bigl(\sum_{n=o}^{\infty}-1^{n_{+}}|n_{+}\rangle \langle n_{+}|
\bigr)|\zeta \rangle\; =\;
    {{(1 - tanh^{2}\zeta)} \over {(1 + tanh^{2}\zeta)}}.
\end{equation}   
Hence, with {\it parity} as the relevant variable the state allows maximal BIQV
 only at $\zeta \rightarrow \infty$ - only at this limit does average 
value (of the relevant dynamical variable for either channel) vanishes.\\
We have shown thus that correlations among polarizations as dynamical 
variables (with expectation values between $-1$ and $1$) allow 
maximal Bell inequality violation for a product of
 two two-mode squeezed vacuum states. The Wigner function of the state 
is non-negative. This same state for
{\it parity} as the dynamical
variable allows maximal violation of Bell's inequality only in 
the limit of the squeezing paprameter  $\zeta \;\; \rightarrow \;\; \infty$, 
 i.e., only 
when the state's expectation value for either parity is nil. Hence we have 
demonstrated that the 
violation incurred in Bell's inequality is strongly related not only to the   
 entanglement (of the state) and the particular dynamical variables 
- i.e. the variables involved in the correlations considered in the Bell 
operator but that maximal entanglement is not necessary for maximal 
Bell's inequality violation.\\  
{\bf Acknowledgement} The research of FK is supported by NSERCC and the 
research of AS is supported by CNPq.\\

\end{document}